\documentstyle[fleqn]{tp}

\input epsf

\def\etal{{\it et\thinspace al\/\thinspace}}

\begin{document}

\twocolumn[
\title{The implications of the large scale galaxy power spectrum for 
       Cold Dark Matter}
\author{C. M. Baugh$^1$ and E. Gazta\~{n}aga$^2$ \\
{\it $^1$Department of Physics, South Road, Durham, DH1 3LE} \\
{\it $^2$Institut d'Estudis Espacials de Catalunya, Barcelona}}
\vspace*{16pt}   

ABSTRACT.\
The APM Galaxy Survey maps the angular positions of more than 
one million galaxies to $b_{J} = 20$, covering a volume comparable 
to the forthcoming Sloan Digital Sky Survey.
A numerical algorithm has been developed to estimate the power 
spectrum in three dimensions, using the angular clustering and 
a model for the redshift distribution of APM Galaxies.
The power spectrum obtained is free from distortions of the 
pattern of clustering caused by the peculiar motions of galaxies.
We discuss the uncertainties in the estimated power spectrum and 
describe tests of the algorithm using large numerical simulations.
The APM Galaxy power spectrum shows an inflection at a wavenumber 
$k \sim 0.15 h {\rm Mpc}^{-1}$, with evidence for a peak or turnover 
in the range $ k \sim 0.03-0.06 h {\rm Mpc}^{-1}$. These features can place 
strong constraints on Cold Dark Matter models for structure formation.
\endabstract]

\markboth{C.M. Baugh \& E. Gazta\~{n}aga}{Galaxy power spectrum on large scales}

\small

\begin{table*}
\caption[]{
The variants of the Cold Dark Matter model used in the comparison 
with the APM Survey power spectrum in Figures 1 and 2. The parameter 
$\Gamma$ describes the shape of the power spectrum. Apart from the 
COBE-CDM model, the models are normalised to match the local abundance of 
rich clusters (Eke etal 1996).
}
  \centering
  \begin{tabular}{l|*{5}{c}}
    \hline
    Model & $\Omega_{0}$ & $\Lambda_{0}$ & $\Gamma$ &
    $\sigma_{8}$ \\
    \hline
    cluster-CDM         & 1.0 & 0.0 & 0.50 & 0.52 \\
    COBE-CDM            & 1.0 & 0.0 & 0.50 & 1.24 \\
    $\tau$CDM            & 1.0 & 0.0 & 0.20 & 0.52 \\
    $\Lambda$CDM        & 0.3 & 0.7 & 0.20 & 0.93 \\
    OCDM                & 0.4 & 0.0 & 0.25 & 0.76 \\
    \hline
  \end{tabular}
\label{tab:1}
\end{table*}

\begin{figure*}
{\epsfxsize=10.5truecm \epsfysize=14.5truecm 
\epsfbox[60 170 560 660]{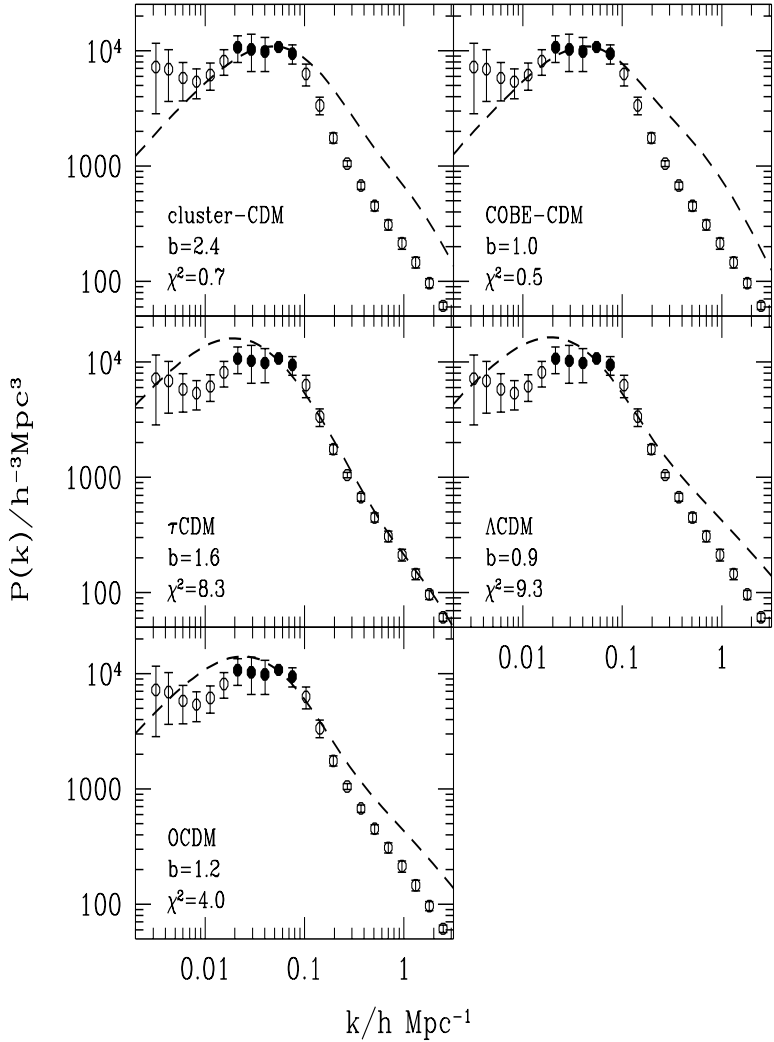}}
\caption[]
{
A comparison of the CDM models listed in Table 1 with the 
APM power spectrum on large scales. 
The APM power spectrum is shown by the circles and the errorbars show 
the $1\sigma$ error on the mean obtained by splitting the APM Survey 
into 4 zones (taken from Table 2 of Gazta\~{n}aga \& Baugh 1998). 
The estimates of the power used in the comparison are indicated by 
the filled circles. The panels show a label indicating the CDM model 
and the value of the bias parameter that give the best match the 
galaxy power spectrum over the indicated range of wavenumbers. 
The values of $\chi^{2}$ are per degree of freedom.
}
\label{fig:1}
\end{figure*}

\begin{figure*}
{\epsfxsize=10.5truecm \epsfysize=14.5truecm 
\epsfbox[60 170 560 660]{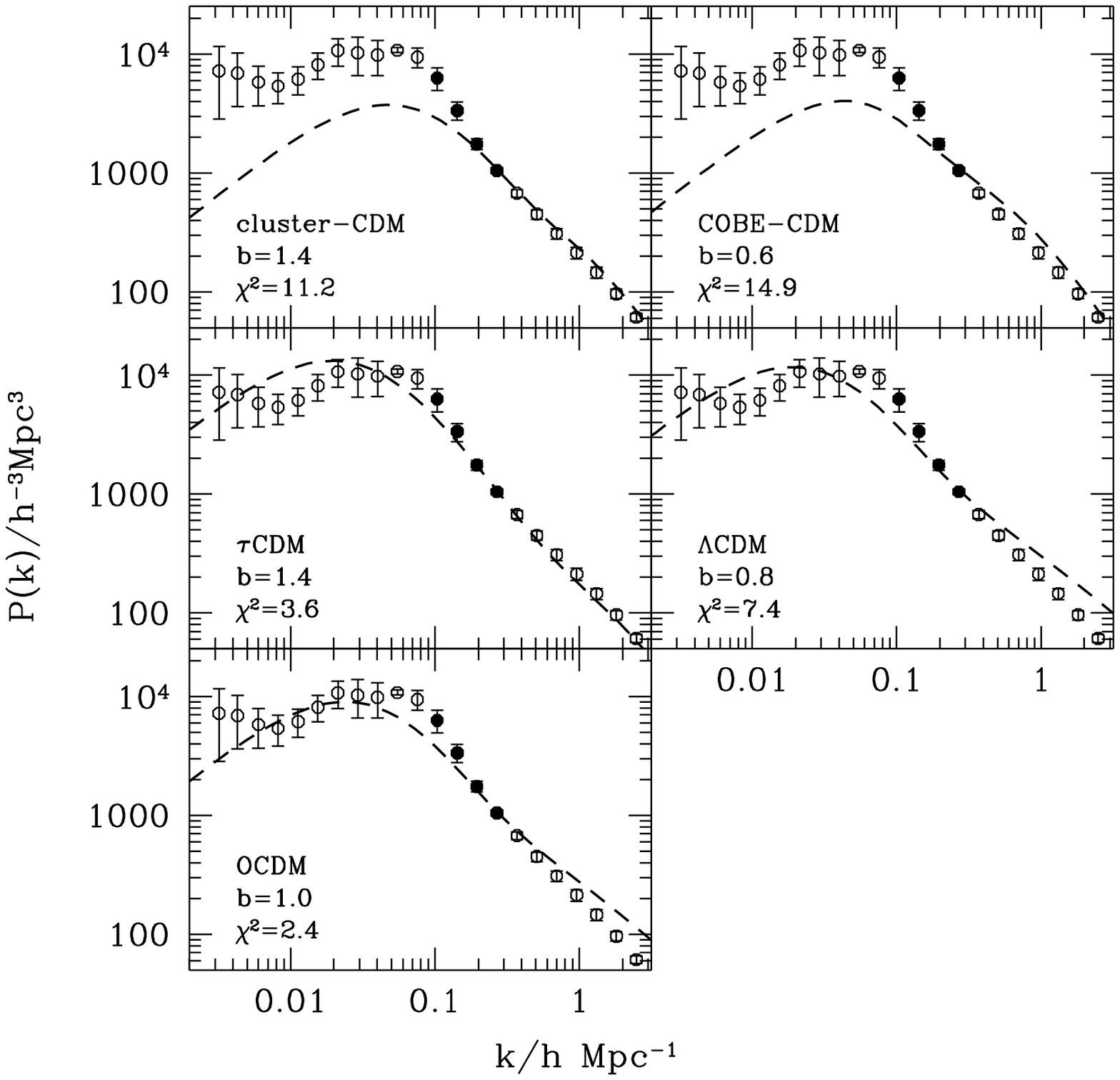}}
\caption[]
{
The models listed in Table 1 are now compared with the shape of the 
APM power spectrum around the inflection point, again the points used in the 
fits are indicated by the filled circles.
}
\label{fig:2}
\end{figure*}

\section{Introduction}

The primordial power spectrum of density fluctuations in the universe 
is a key ingredient of any model for structure formation. Unfortunately, 
there are many effects that prevent a direct measurement of this 
quantity. 
Firstly, the shape of the mass power spectrum changes with redshift when the 
{\it rms} fluctuations on a particular scale approach unity and  
couple to density perturbations on other length scales  
(e.g. Baugh \& Efstathiou 1994a; Peacock \& Dodds 1994). 
Secondly, structures are traced out by galaxies and the relation 
between these objects and the underlying density field is 
complex and may vary with redshift (Kauffmann etal 1998).
An illustration of this is provided by the so-called Lyman break galaxies
that have been used to probe structure in the universe at high redshift, 
$z \sim 3$ (see the contribution of C. Steidel 
to these proceedings).
The clustering amplitude of Lyman break galaxies is similar to 
that of bright present day galaxies (Adelberger etal 1998; 
Giavalisco etal 1998).
However, in any viable hierarchical model for structure formation, 
the clustering amplitude of the dark matter at this epoch is much 
lower than this. 
The Lyman-break galaxies are therefore {\it biased} tracers of the 
mass distribution.
Finally, when the redshift of a galaxy is used to infer its spatial 
position, the pattern of clustering is distorted due to the peculiar 
motion of the galaxy resulting from inhomogeneities in its local 
gravitational field. 

In view of these problems, there would appear to be little hope of 
learning anything about the distribution of dark matter in the 
universe by measuring galaxy clustering. 
However, on large scales, $\lambda > 20 h^{-1} {\rm Mpc}$, 
the above phenomena are responsible for introducing differences between 
the power spectra of galaxies and of mass that are either small in magnitude 
or which can for the most part be accurately modelled.

The forthcoming Anglo Australian 2dF and the Sloan Digitial Sky 
Surveys will contain an order of magnitude more galaxy redshifts than 
the largest currently available surveys 
(see the contribution of A. Szalay to these proceedings). 
One result that will emerge from these surveys will be an accurate 
measurement of the power spectrum of galaxy clustering on large scales.
In this article, we discuss how the parent catalogue for the 2dF Survey, 
the APM Galaxy Survey (Maddox etal 1996), 
can provide a  measurement of the power spectrum on comparably large scales.

The APM Survey contains more than one million galaxies down to a magnitude 
limit of $b_J= 20$ and covers $4300$ square degrees. The volume covered by 
the angular APM catalogue is comparable to the volume that the Sloan 
Survey will probe.
Baugh \& Efstathiou (1993 -- BE93; 1994b) developed a numerical algorithm to 
iteratively deproject the angular clustering of galaxies to measure 
the three dimensional galaxy power spectrum.
The power spectrum is estimated in a series of wavenumber bins; 
hence the recovered spectrum does not rely on the 
assumption of a particular parametric form.
Furthermore, the power spectrum  obtained is free from the 
distortion of the pattern of clustering caused by the peculiar 
motions of galaxies, which affects the power spectrum measured 
from redshift surveys.
In this article we review some of the uncertainties involved in the 
deprojection process and discuss the implications of the shape of the 
APM power spectrum for the Cold Dark Matter family of structure 
formation models.

\section{Uncertainties in the Deprojection Algorithm}

The power spectrum in three dimensions is estimated by numerically 
inverting Limber's  equation, which equates the angular correlation 
function to an integral over the spatial two point correlation function 
or power spectrum, and the redshift distribution of galaxies.
Below we summarize some of the areas that introduce uncertainties into the 
the estimate of the power spectrum.
Full details of the deprojection algorithm and tests of accuracy and 
convergence can be found in BE93.  
The algorithm is tested against synthetic APM Survey maps made from 
numerical simulations by Gazta\~{n}aga \& Baugh (1998).

\subsection{Cosmology}
The angular separation versus coordinate distance relation depends upon the 
choice of background cosmology. BE93 investigated the 
effects of different assumptions for the value of the density parameter 
$\Omega$ on the estimated power spectrum. There is no change in the shape 
of the power spectrum for different values of $\Omega$, but a small change in 
amplitude, of around $15$\%.

\subsection{Galaxy Redshift Distribution}
BE93 gave a simple parametric form for the redshift 
distribution of APM Survey galaxies.
This was subsequently found to be in good agreement with the 
redshift distribution of galaxies in the 2dF Survey (see S. Maddox in this 
volume).
Again, uncertainties in the median redshift of APM galaxies do not alter 
the shape of the recovered power spectrum, though can affect the amplitude 
by approximately $5\%$.

\subsection{Approximations involved in Limber's equation}

The derivation of Limber's equation assumes that the depth 
of the survey is much greater than the scale of any clustering in the 
survey. The largest scales on which the APM power spectrum is 
recovered are a significant fraction of the depth of the survey.
Gazta\~{n}aga \& Baugh (1998) present tests of the 
deprojection algorithm using mock APM catalogues 
constructed by the projection of large N-body simulations.
An accurate recovery of the turnover in the three dimensional power 
spectrum is possible in simulation boxes of side $600h^{-1}$Mpc, which 
is smaller than the radial extent of the APM Survey.

\subsection{Evolution of galaxy clustering}

The evolution of galaxy clustering is a complex interplay between 
the growth of fluctuations in the dark matter and galaxy formation. 
Fluctuations on a particular scale grow at a rate different to that 
predicted by linear perturbation theory when the {\it rms} variance 
approaches unity.
At higher redshifts, the galaxy population sampled in the APM Survey 
becomes brighter and it is possible that these galaxies have different 
intrinsic clustering properties to those seen at lower redshift, though 
there is little evidence for this being a strong effect in this Survey 
(Maddox \etal 1996; see also Tadros \& Efstathiou 1996 and  Hoyle \etal 1998).
The question of the evolution of the bias of APM galaxies can now be 
addressed using semianalytic models for galaxy formation (e.g. 
Benson \etal 1998).
In view of the low median redshift, $z\sim 0.13$ of APM galaxies to 
$b_J \sim 20$, both of these effects are expected to be small. 
For this reason, BE93 used the simplest approximation that 
$P(k,z) = P(k)/(1+z)^{\alpha}$ and adopted the case where the power spectrum 
is fixed in comoving coordinates, $\alpha=0$. 
Maddox, Efstathiou \& Sutherland (1996) adopt a value of $\alpha=1.3$, 
which corresponds to clustering fixed in proper coordinates. 
Changing the value of $\alpha$ has essentially no effect on the shape of 
the power spectrum, and has only a small effect on the amplitude 
that is recovered, with the range of interesting values of 
$\alpha$ translating into a $20$\% uncertainty in the amplitude.

\section{The Implications for Cold Dark Matter}

We compare the shape and amplitude of the APM power spectrum 
with popular variants of the Cold Dark Matter (CDM) model in 
Figures 1 and 2. The model parameters are listed in Table 1. 
The nonlinear form of the CDM power spectrum, calculated using the 
formula given by Peacock \& Dodds (1996) is shown by the dashed lines 
in the figures. One further degree of freedom, a {\it scale independent}  
bias is allowed. The APM power spectrum shows evidence for a turnover or 
flattening in slope on scales around $k\sim 0.03-0.06 h{\rm Mpc}^{-1}$ and 
an inflection at $k\sim 0.15 h{\rm Mpc}^{-1}$.
We consider the best fit for the models to the turnover 
in the power spectrum in Figure 1 and to the inflection in Figure 2. 
In each panel, the best fitting value of the bias parameter is given, 
along with the corresponding value of $\chi^{2}$ per degree of freedom.

The CDM models with a shape parameter denoted by $\Gamma$ of 
$\Gamma=0.5$ do the best out of the examples considered at matching the 
position of the turnover.
These models are however the worst at matching the inflection point, 
which models with  a lower value of $\Gamma$ reproduce better.
Thus none of the CDM models considered does particularly well at matching 
the location of the turnover and the shape of the power spectrum at the 
inflection point simultaneously (see also the analysis of 
Gawiser \& Silk 1998). 

The situation could be improved if the assumption that bias is 
independent of scale is dropped. The form of the effective bias is then 
is given by $b_{eff}=(P_{gal}(k)/P_{mass}(k))^{1/2}$. 
The effective bias required is a nonmonotonic 
function of scale in all cases. 
For a CDM universe with a cosmological 
constant, the bias varies between $b=0.74 \pm 0.13$ at $k=0.03h$Mpc$^{-1}$ to 
$b=1.00 \pm 0.11$ at $k=0.1h$Mpc$^{-1}$,  
before becoming an {\it antibias} again of 
$b=0.75 \pm 0.02$ at $k=0.3h$Mpc$^{-1}$.
Whilst a constant bias is undoubtably a poor approximation on scales of a few 
megaparsecs and smaller (Benson \etal 1998), the above  degree of nonmonotonic 
change in the bias parameter on scales of tens to hundreds of megaparsecs 
would appear difficult to motivate physically.

\end{document}